\documentstyle[11pt,aaspp4]{article}
\def\be{\begin{equation}}

\def\ee{\end{equation}}

\def\etal{et al.}

\def\hMpc{h^{-1}{\rm Mpc}}
\def\h3Mpc{h^{-3}{\rm Mpc}^3}
\def\hcMpc{h^{-3}{\rm Mpc}^3}
\def\h3Mpcinv{h^{3}{\rm Mpc}^{-3}}

\def\ten#1{\times 10^{#1}}

\def\refer#1{#1}
\def\refeq#1{\relax}		% for 'quiet' ref
\def\IAUOC{in IAU Symp. 124, Observational Cosmology}
\def\IAU130{in IAU Symp. 130, Large Scale Structures of the Universe}

\slugcomment{EFI-97-11.  Revised May 19, 1997}

\begin{document}

\title{The local space density of dwarf galaxies}

\author{Jon Loveday}
\affil{Department of Astronomy and Astrophysics, University of Chicago, 
5640 S. Ellis Ave, Chicago, IL 60637, USA;
	loveday@oddjob.uchicago.edu}

\begin{abstract}
We estimate the luminosity function of field galaxies over a range of ten 
magnitudes ($-22 < M_{B_J} < -12$ for $H_0 = 100$ km/s/Mpc)
by counting the number of faint APM galaxies around Stromlo-APM redshift
survey galaxies at known distance.
The faint end of the luminosity function rises steeply at 
$M_{B_J} \approx -15$, implying that the space density of dwarf galaxies is
at least two times larger than predicted by a
Schechter function with flat faint-end slope.
Such a high abundance of dwarf galaxies at low redshift can help 
explain the observed number counts and redshift distributions of
faint galaxies without invoking exotic models for galaxy evolution.
\end{abstract}

\keywords{galaxies: distances and redshifts
--- galaxies: evolution
--- galaxies: luminosity function, mass function
--- cosmology: observations
--- surveys}

\section{Introduction}

Knowledge of the space density of dwarf galaxies in the local Universe is
of great importance both in its own right, for example to constrain
models of galaxy formation and to understand the local distribution of
matter, and also in order to interpret observations of faint galaxies.
For quite some time now (see, for example, the review by 
\refer{Koo and Kron 1992}), galaxy number counts, particularly
 in blue passbands, have been
found to increase faster with apparent magnitude than predicted by
simple no-evolution models, whereas the redshift distribution of
galaxies in faint surveys {\em is} compatible with no evolution.
Various models have been proposed to account for this discrepancy,
including a disappearing or fading population of dwarf galaxies
(\refer{Broadhurst \etal\ 1988}, \refer{Cowie \etal\ 1991}, 
\refer{Babul and Rees 1992}), non-conservation of galaxy numbers through
merging (\refer{White 1989}, \refer{Cowie \etal\ 1991}, 
\refer{Broadhurst \etal\ 1992}) or adoption of a non-zero cosmological constant
(\refer{Fukugita \etal\ 1990}).
The Automated Plate Measurement (APM) galaxy number counts of
\refer{Maddox \etal\ (1990c)} show a factor of two higher counts at $b_J = 20.5$
than expected from a no-evolution model, thus further exacerbating the
discrepancy between observed number counts and redshift distributions.

Other authors, however, (\refer{Koo and Kron 1992}, \refer{Koo \etal\ 1993}, 
\refer{Gronwall and Koo 1995}), have suggested that modifying the assumptions
going into the no-evolution models can help reconcile the observed 
number counts and redshift distributions, without resorting to exotic
evolution models or non-standard cosmologies.
In particular, no-evolution models have traditionally assumed a
\refer{Schechter (1976)} form for the luminosity function with moderate 
faint-end slope, eg. $\alpha = -1.25$ (\refer{Ellis 1987}) or
$\alpha = -1.15$ (\refer{Lilly 1993}).
\refer{Gronwall and Koo (1995)} show that number counts in $K$, $R$ and $B_J$
bands, as well as colour and redshift distributions, are matched well
by a no-evolution model in which the faint end of the galaxy luminosity 
function ($M_{B_J} \gtrsim -15$\footnote{Throughout, we assume a 
Hubble constant $H_0$ of 100 km/s/Mpc.}) 
is dominated by blue galaxies ($B - V \le 0.6$)
and rises significantly above a Schechter function with flat faint-end slope.
To date, there have been few measurements of the shape of the local 
field galaxy luminosity function at such low luminosities.
\refer{Eales (1993)} measured an apparent upturn in the luminosity function 
(LF) at $M_{B_J} \gtrsim -15$ for galaxies at $z < 0.1$.
However, he used the $1/V_{\rm max}$ method to compute the LF, and so it
is subject to bias due to an inhomogeneous distribution of a very small
number of galaxies.
A similar caution should be applied to the results of
\refer{Lonsdale and Chokshi (1993)}, whose lowest-luminosity measurement
is anyway consistent with a flat faint-end Schechter function due to
Poisson statistics alone.
The analysis by \refer{Marzke \etal\ (1994)} of the CfA Redshift
Survey shows evidence for an upturn in the LF at low luminosities,
although the possibility of a scale error in the Zwicky magnitude
system makes the amplitude of the faint-end excess uncertain
(but see \refer{Takamiya and Kron 1995}).
Probably the best evidence for an upturn in the LF at low luminosities
comes from a recent measurement of the galaxy luminosity function
from the ESO Slice Project (\refer{Zucca \etal\ 1997}), who find
an excess of galaxies fainter than $M_{B_J} \approx -17$ above their
best-fit Schechter function.

In this paper we push the measurement of the field galaxy luminosity function
to fainter limits than previous work 
by counting the statistical excess of faint
($b_J < 20.5$) galaxies in the APM Galaxy Survey seen in projection around
local galaxies at known distance in the Stromlo-APM Redshift Survey.
The method by which we estimate the luminosity function is described
in the following section.
The galaxy samples and the application of the method are described in 
\S\ref{sec:samples} and in \S\ref{sec:just} we justify the use of nearby
centre galaxies to measure the faint-end of the galaxy LF.
Results are presented in \S\ref{sec:results} and several tests of
our procedure are described in \S\ref{sec:tests}.
We conclude in \S\ref{sec:concs}.

\section{Method} \label{sec:method}

Given a catalogue containing positions and magnitudes for a large number
of galaxies of unknown redshift and a smaller redshift survey 
in the same area of sky, one can measure the
galaxy luminosity function to fainter luminosities than using the redshift
survey alone.
The technique used here to measure the abundance of dwarf galaxies
relies on the assumption that correlated galaxies seen close in projection
on the sky to a galaxy of known distance are also at the same distance.
Although we do not know individually which galaxies are correlated,
we can statistically determine the numbers of associated galaxies as a
function of apparent magnitude, and hence absolute magnitude.
Repeating this for a large number of centre galaxies, we can measure
a luminosity function with good statistical accuracy at the faint end.

Perhaps the most straightforward approach is to count the excess number
of galaxies in bins of absolute magnitude out to a fixed projected
separation around each centre galaxy, and to sum over centre galaxies.
This approach was employed by \refer{Phillipps and Shanks (1987)}, who counted
the numbers of galaxies down to $b_J = 20.5$ from COSMOS scans of 
United Kingdom Schmidt Telescope (UKST) plates
about centre galaxies taken from various Durham redshift surveys,
to measure the field galaxy LF over the magnitude range 
$-20.5 < M_{B_J} < -16$.
However, as pointed out by \refer{Saunders \etal\ (1992)}
in a slightly different context, such an approach is not very efficient
as centre galaxies at small distances contribute little signal but a large
number of galaxies to the sum, since
nearby a fixed projected separation corresponds to a large solid angle,
and so the excess neighbours are dominated by projection effects.
For this reason, Phillipps and Shanks only used centre galaxies at
distances larger than $100 \hMpc$, thus limiting their ability to
probe the faint end of the LF.

Here, we generalize the method of 
\refer{Saunders \etal\ (1992)} to estimate a (noisy)
LF for each centre galaxy and then combine the LF estimates in a
minimum-variance way.
Saunders \etal\ counted galaxies in bins of projected separation $\sigma$
about centre galaxies of known redshift to estimate the projected
correlation function, $\Xi(\sigma)$, which is related to the spatial
correlation function, $\xi(r)$, by
\be 
\Xi(\sigma) = \int_{-\infty}^{+\infty} 
\xi( \sqrt{ \Delta y^2 + \sigma ^2 } ) d \Delta y.
\ee
We count galaxies as a function of $\sigma$ and magnitude
to estimate the quantity $X(M,\sigma) = \phi(M) \Xi(\sigma)$.
Assuming a model for $\Xi(\sigma)$ then yields the luminosity function 
$\phi(M)$.

\subsection{Estimating $X(M,\sigma)$ from a single centre galaxy}

Consider a centre galaxy at known distance $y$.
We count the number of galaxies $n$ in bins of projected separation 
$\sigma \pm \delta\sigma/2$ and apparent magnitude $m \pm \delta m/2$ 
about this centre galaxy.
The expected value of $n$ is
\be
\langle n \rangle = \frac{2 \pi a \sin \theta \delta\sigma}{y} 
	\int_0^\infty \phi[M(m,x)] \delta m\, x^2 [1 + \xi(r)] dx
\label{eqn:nexact}
\ee
where $r^2 = x^2 + y^2 - 2xy \cos(\theta)$ is the square of the 
physical separation between galaxies at
distance $x$ and $y$ with angular separation $\theta = \sigma/y$,
$a$ is the fraction of the projected annulus within the survey boundary and 
$\phi(M) \delta m$ is the 
number density of galaxies of absolute magnitude $M \pm \delta m/2$.
The absolute magnitude $M$ as a function of apparent magnitude $m$
and distance $x$ (in $\hMpc$, with corresponding redshift $z$) 
is given by the usual formula:
$M(m,x) = m - 5\lg [x(1+z)] - 25 - 3z$,
where the final $-3z$ term is an approximate 
$k$-correction for galaxies of unknown type in the $b_J$ passband.

For centre galaxies at moderate distance $y$ 
($y \gg \sigma$ and $\xi(y) \ll 1$),
\be
\langle n \rangle \approx \frac{2 \pi a \sin \theta \delta\sigma}{y} 
	\left[\bar{N}(m)\delta m + 
	\phi[M(m,y)]\delta m\, y^2 \Xi(\sigma) \right],
\ee
where $\bar{N}(m)\delta m$ is the surface density of galaxies of apparent 
magnitude $m \pm \delta m/2$ in the 2-d catalogue.

As discussed by Saunders \etal, this approximation is subject to biases
due to the fact that for nearby galaxies, $x^2$ may grow faster than
$\xi(r)$ falls off, and so excess pairs may not be close to the galaxy
of known redshift.
Conversely, at large distances, the selection function may be falling
off so steeply that neighbours in projection will on average tend to be
nearer than $y$.
These biases are corrected for as discussed below.

To correct for the survey boundary, we count the number $n_r$ of randomly 
distributed points of mean surface density $\bar{N}_r$
in the same $\sigma$ bins.
Scaling $n_r$ to the surface density of galaxies of magnitude $m$,
$n'(m, \sigma) = \frac{\bar{N}(m)}{\bar{N}_r} n_r(\sigma)$, we expect
\be
\langle n' \rangle = \frac{2 \pi a \sin \theta \delta \sigma}{y} 
\bar{N}(m)\delta m.
\ee
Our estimate of the relative excess $X(M,\sigma) = \phi(M) \Xi(\sigma)$ is then
\be
X(M,\sigma) = \frac{1}{p(m,\sigma,y)} \left[\frac{n}{n'} - 1 \right],
\ee
where
\be
p(m,\sigma,y) = \frac{1}{\bar{N}(m)} k(m,\sigma,y) y^2
\ee
corrects for projection effects and the factor $k(m,\sigma,y)$ corrects
for biases caused by the small angle/large distance approximation.
From equation (\ref{eqn:nexact}), the expectation value of $X$ is
\be
\langle X(M,\sigma) \rangle = \frac{1}{y^2 k(m,\sigma,y)} 
	\int_0^\infty \phi[M(m,x)] x^2 \xi(r) dx
\ee
and so setting
\be
k(m,\sigma,y) = \frac{\int_0^\infty \phi[M(m,x)] x^2 \xi(r) dx}
	             {y^2 \Xi(\sigma) \phi[M(m,y)]}
\ee
will make $X(M,\sigma)$ an unbiased estimate of $\phi(M) \Xi(\sigma)$.
Our estimator for $\phi(M)$ thus depends on $\phi(M)$.
However if $\phi(M)$ is modeled  as  a smooth function (eg. a Schechter
function), a stable solution is reached within ten iterations.

\subsection{Combining $X(M,\sigma)$ estimates to obtain $\phi(M)$}

Each $X_i(M,\sigma)$ calculated above for each centre galaxy should be an
unbiased, albeit noisy, estimator of the quantity $\phi(M) \Xi(\sigma)$.
We now wish to combine these estimates in an optimal way
to obtain a close to minimum variance estimate of $\phi(M)$.

Again following Saunders et al, the expected variance in the quantity $n_i$ is
\be
{\rm Var}(n_i) \approx n'_i [1 + p(m,\sigma,y) \phi(M) \Xi(\sigma)]
			    [1 + \bar{N}(m) J_2(\theta)]
			    [1 + f \phi(M) J_3(\sigma)],
\ee
where
\be 
J_2(\theta) = 2 \pi \int_0^\theta \theta' w(\theta') d\theta',
\ee
and 
\be
J_3(\sigma) = 4 \pi \int_0^\sigma \sigma'^2 \xi(\sigma') d\sigma',
\ee
and $f$ is the total fraction of galaxies with measured redshifts.
In combining the individual measurements $X_i$ to an overall estimate for
$\phi(M)$, we sum over centre galaxies, weighting 
each measurement by $b_i = [\Xi(\sigma)]^2/{\rm Var}(X_i)$,
\be
\phi(M) = \frac{\sum_i b_i X_i/\Xi(\sigma)}{\sum_i b_i},
\label{eqn:phi_wt}
\ee
\be
\frac{1}{b_i} = \frac{1}{n'_i [p(m,\sigma,y) \Xi(\sigma)]^2} 
		[1 + p(m,\sigma,y) \phi(M) \Xi(\sigma)]
		[1 + \bar{N}(m) J_2(\theta)]
		[1 + f \phi(M) J_3(\sigma)].
\ee
The expected variance on our final measure of $\phi(M)$ is then
\be
{\rm Var}[\phi(M)] = 1/\sum_i b_i.
\label{eqn:phi_errs}
\ee
%The optimal weighting of the $X_i$ also depends on an assumed $\phi(M)$,
%and so we calculate new $X_i$ and an overall $\phi(M)$ in each iteration.

\section{Galaxy Samples and Application} \label{sec:samples}

For centre galaxies, we use the Stromlo-APM Redshift Survey
(\refer{Loveday \etal\ 1996}), which consists of 1787 galaxies with
$b_J \le 17.15$ selected randomly at a rate of 1 in 20 from the APM
Galaxy Survey (Maddox \etal\ 1990a,b).
\refeq{Maddox \etal\ 1990a}\refeq{Maddox \etal\ 1990b}
We count APM galaxies around each redshift survey galaxy in five linearly
spaced bins of projected separation $\sigma$ up to $5 \hMpc$ and in
22 evenly spaced bins in apparent magnitude from $b_J = 15$ to 20.5.
There are a total of 2,389,032 APM galaxies in this magnitude range.
In order to correct for survey boundaries and holes drilled out around bright 
stars and large galaxies, 
we use a random catalogue of 3,476,477 random points uniformly distributed
over the area of the APM survey.
Although there are only about 1.5 times as many random points as galaxies 
overall, the number
of random points is significantly larger than the number of galaxies (527,611)
in the faintest apparent magnitude bin.
Thus shot noise in the random catalogue introduces negligible random
error into our results.

When counting galaxy and random points in projection around centre galaxies,
we avoid objects within one major diameter of the centre galaxy.
This is to avoid three biases that might otherwise be caused by a large 
foreground 
galaxy: (1) obscuration of faint galaxies directly behind it, (2) lensing
of background galaxies and (3) break-up of a large
galaxy by the APM machine into spurious sub-images, artificially boosting
the apparent number of projected neighbours.
The worst of these latter cases have already been 
removed by the ``holes'' drilled around large images during the construction
of the APM survey (\refer{Loveday 1996}, Maddox \etal\ 1990a).
The majority of galaxies in the Stromlo-APM survey have $\mu_{b_J} \approx 25$
isophotal major diameters of 20-100 arcsec.

We assume a power-law form for the spatial correlation function,
$\xi(r) = (r/r_0)^{-\gamma}$, with $\gamma = 1.71$ and $r_0 = 5.1 \hMpc$
(\refer{Loveday \etal\ 1995}).
The projected correlation function is then given by
\be
\Xi(\sigma) = (\sigma/\sigma_0)^{1-\gamma},
\ee
where
\be
\sigma_0 = \left[r_0^\gamma\, \Gamma\!\left(\frac{1}{2}\right)
	   \Gamma\!\left(\frac{\gamma-1}{2}\right)/
	   \Gamma\!\left(\frac{\gamma}{2}\right)\right]^{\frac{1}{\gamma-1}}.
\ee

For an initial luminosity function $\phi(M)$, we use the Schechter
function fit from Loveday \etal\ (1992).
We store the $X_i$ calculated for each centre galaxy by
projected separation bin and {\em apparent} magnitude bin in order to avoid
incompleteness effects which would otherwise result if we used {\em absolute}
magnitude bins.
We convert apparent to absolute magnitudes when we combine the $X_i$
estimates in equation~(\ref{eqn:phi_wt}), using 22 bins in absolute
magnitude from $M = -22$ to $M = -11$.
At this stage, we also have the choice of which $\sigma$-bins to use;
for example we can count galaxies at projected separation 0--1,
0--2 or 2--5 $\hMpc$.
The consistency of the estimates $\phi(M)$ for a range of projected separations
will provide important confirmation of our results.

\section{Justifying the Use of Nearby Centre Galaxies} \label{sec:just}

Before we present the results of this analysis, we first justify
our assumption that excess faint galaxies seen close in projection
to nearby redshift survey galaxies really are physically correlated.
The APM galaxy survey has a magnitude limit of $b_J = 20.5$,
corresponding to a limiting depth for $L^*$ galaxies of $\sim 600 \hMpc$,
whereas galaxies of absolute magnitude $M_{B_J} = -15$ can only
be seen out to a distance of $115 \hMpc$.
One might therefore wonder whether fluctuations in the number of 
uncorrelated background galaxies might swamp the contribution from
genuinely correlated dwarf galaxies for nearby centres.
We therefore perform a simple experiment to show that the number
of APM galaxies counted to various magnitude limits is systematically
larger close to a nearby centre galaxy than in a random control field.

Taking in turn those $N_{\rm cen} = 477$ galaxies in the Stromlo-APM Survey 
that are closer than $100 \hMpc$ as centres, we count the number of APM
galaxies ($N_{\rm gal}$) within a projected separation of $1 \hMpc$ 
and also the number of background galaxies ($N_{\rm bgr}$)
within the same solid angle at a randomly chosen location.
We use the random catalogue to correct for solid angle lost due to
the survey boundary and holes drilled out around large images.
This is done for APM galaxies to a magnitude limit of $b_J = 20.0$, 19.0, 18.0,
and 17.0.
In figure~\ref{fig:bgrtest}, we plot the frequency histograms
(over the 477 centre galaxies) of the ratio $N_{\rm gal}/N_{\rm bgr}$
(solid histogram) and its inverse $N_{\rm bgr}/N_{\rm gal}$ (dotted histogram).
One can clearly see that the ratio $N_{\rm gal}/N_{\rm bgr}$ is systematically
greater than unity, ie. there is a statistical excess of APM galaxies
near centre galaxies.
One can quantify this excess by measuring the mean and rms of the
ratio $N_{\rm gal}/N_{\rm bgr}$ over centres.
The mean and its standard error
[$\sigma(N_{\rm gal}/N_{\rm bgr})/\sqrt{N_{\rm cen}}$]
are shown in Table~\ref{tab:bgrtest}.
Also shown in this table is the probability from the Kolmogorov test
that the quantities $N_{\rm gal}$ and $N_{\rm bgr}$ come from the same
distribution.
We see, particularly for the brighter magnitude cuts, that the 
distributions $N_{\rm gal}$ and $N_{\rm bgr}$ are significantly different,
and that the ratio $N_{\rm gal}/N_{\rm bgr}$ is systematically larger
than unity at at least the 4-$\sigma$ level.
From this test we conclude that we are justified in assuming that
the excess APM galaxies seen close to Stromlo-APM centre galaxies
are physically correlated, even for centre galaxies much closer than
the characteristic depth of the APM galaxy survey.

\begin{table}[htbp]
 \begin{center}
 \caption{Galaxy counts about centre galaxies $N_{\rm gal}$ and control fields
	  $N_{\rm bgr}$}
 \vspace{0.5cm}
 \label{tab:bgrtest}
 \begin{math}
 \begin{array}{ccc}
 \hline
 \hline
 m_{\rm lim}^a & \langle N_{\rm gal}/N_{\rm bgr} \rangle^b & 
\mbox{K-S Test}^c \\
 \hline
 a\ 20.0 & 1.08 \pm 0.017 & 2.3 \ten{-1} \\
 b\ 19.0 & 1.16 \pm 0.025 & 4.9 \ten{-2} \\
 c\ 18.0 & 1.35 \pm 0.041 & 2.2 \ten{-3} \\
 d\ 17.0 & 1.90 \pm 0.109 & 2.1 \ten{-6} \\
  \hline
  \hline
 \end{array}
 \end{math}
\end{center}

$^a$ Magnitude limit of APM galaxies.\\
$^b$ Mean and standard error in the ratio $N_{\rm gal}/N_{\rm bgr}$.\\
$^c$ Probability that $N_{\rm gal}$ and $N_{\rm bgr}$ are drawn from
the same distribution.
\end{table}

\section{Results} \label{sec:results}

In figure~\ref{fig:phi}(a) the points with error bars show
the first-iteration estimate of the luminosity
function $\phi(M)$, counting galaxies to projected separation 
$\sigma \le 1 \hMpc$.
For comparison, this plot also shows (as the continuous line) our
earlier Schechter function fit to $\phi(M)$ using the redshift survey 
galaxies alone (\refer{Loveday \etal\ 1992}) plus its extrapolation
to lower luminosities (dotted line).
Comparing with the earlier result, we see three regimes.
(1) For $M \lesssim -18$, we see good agreement with the Schechter function.
(2) For $-18 \lesssim M \lesssim -15$ we see that the new determination of 
$\phi(M)$ lies $\sim 2\sigma$ {\em below} the Schechter function.
(3) At the faintest luminosities, $M \gtrsim -15$, we see a sharp rise
in $\phi(M)$, rising well above the Schechter function.
Thus there appears to be a significant excess of faint galaxies above
a flat faint-end power-law, but we first need to understand why we
disagree with our earlier estimate of $\phi(M)$ over the magnitude range
$-18 \lesssim M \lesssim -15$.

We believe that this apparent discrepancy is due to the
fact that we are really estimating the product $\phi(M) \Xi(\sigma)$
and we are assuming that $\Xi(\sigma)$, and therefore $\xi(r)$, is fixed,
ie. {\em independent of galaxy luminosity}.
In fact, as we showed in an earlier paper (\refer{Loveday \etal\ 1995}),
there is evidence that low-luminosity galaxies are less strongly
clustered than $L^*$ ($M^* \approx -19.7$) galaxies, 
and thus we would expect our estimate of
$\phi(M) \Xi(\sigma)$ to be biased low at low luminosities, as we indeed 
observe in figure~\ref{fig:phi}(a) over the magnitude range 
$-18 \lesssim M \lesssim -15$.
This effect was also noted by \refer{Lorrimer \etal\ (1994)} in their
analysis of the distribution of satellite galaxies.
The variation of clustering strength with luminosity is too poorly known to
attempt to correct the $\phi(M)$ estimate shown here.
In fact, the present data provides probably our best constraints on how
$\xi(r)$ varies with luminosity.

By absolute magnitude $M \approx -14$ another effect is
dominating: either the luminosity segregation reverses (such that
$M \gtrsim -14$ galaxies are {\em more} strongly clustered about $L^*$ galaxies
than other $L^*$ galaxies)
and/or the intrinsic space density of dwarf galaxies is significantly
higher than predicted by extrapolation of a flat faint-end Schechter function.
As we will see below, the latter seems a more likely explanation for the
observed excess of faint galaxies near $\sim L^*$ galaxies.
It also has the attractive feature of helping to explain the observed 
steep number
counts of faint galaxies compared with traditional no-evolution models.

Figure~\ref{fig:phi}(a) showed our estimate of $\phi(M)$ after just
one iteration, ie. assuming that the true luminosity function is a flat
faint-end Schechter function.
Thus the steep faint-end is not due to an instability in our iteration
procedure.
Figure~\ref{fig:phi}(b) shows our estimate of $\phi(M)$ after
ten iterations, by which time the solution has converged.
We see that the faint-end slope has steepened slightly further.
To model the observed $\phi(M)$, we have fitted a modified form of
the Schechter function, with an additional faint-end power law:
\be
\phi(L) = \phi^* \left(\frac{L}{L^*}\right)^\alpha 
	    \exp\left(\frac{-L}{L^*}\right)
	    \left[1 + \left(\frac{L}{L_t}\right)^\beta\right].
\label{eqn:dpschec}
\ee
In this formulation $\phi^*$, $L^*$ and $\alpha$ are the standard Schechter
parameters, $L_t$ is a transition luminosity between the two power-laws
and $\beta$ is the power-law slope of the very faint-end.
No physical interpretation is intended by this choice of formula,
it is merely a convenient way of modeling the observed $\phi(L)$
over this extended range of luminosity and for estimating the faint-end
slope.
The line in figure~\ref{fig:phi}(b) shows our best-fit
``double power-law'' luminosity function, which has parameters:
$\alpha = -0.94$, $M^* = -19.65$, $\phi^* = 1.54\ten{-2} \hcMpc$, 
$M_t = -14.07$ and $\beta = -2.82$.
Although this fit is poor over the range $-17 \lesssim M \lesssim -14$,
our $\phi(M)$ estimate is almost certainly biased low
over this range by the weaker clustering of galaxies fainter than $L^*$.
Clearly, the faint-end slope $\beta = -2.82$ cannot extend to indefinitely
low luminosities, but it shows no obvious signs of flattening brightward
of $M = -12$.

\section{Checks and Tests} \label{sec:tests}

%In this section we present a number of tests of the reliability
%of the results presented above.

\subsection{Varying projected separation $\sigma$}

In the preceding section we presented results based only on counts of
neighbouring galaxies at a projected separation of $\sigma \le 1 \hMpc$.
By increasing $\sigma$ one decreases shot-noise errors by increasing the
total number of neighbours, although now a smaller fraction of these
neighbours will be genuinely correlated with the centre galaxy.
Figure~\ref{fig:sigtest} shows our estimate of $\phi(M)$ if we count
galaxies to a projected separation $\sigma \le 2 \hMpc$ (solid symbols)
and $\sigma \le 5 \hMpc$ (open symbols).
We see that the estimated $\phi(M)$ is insensitive to the limiting
$\sigma$ used, except that the deficit at $M \approx -15$ worsens as the
maximum projected separation increases, and the error bars increase.
This is to be expected from our earlier results 
on luminosity segregation, since figure~6 of Loveday \etal\ (1995)
shows that the weaker clustering of sub-$L^*$ galaxies is more
pronounced at scales larger than $1 \hMpc$.
The error bars also increase if we decrease the limiting projected 
separation to values of less than $1\hMpc$.
Counting neighbours to $\sigma \approx 1 \hMpc$, as shown in 
Figure~\ref{fig:phi}, thus appears to be about optimal for this analysis.

\subsection{Varying limiting apparent magnitude}

The second test we have performed is to cut back on the magnitude limit
to which we count APM galaxies.
There are two reasons for doing this.
The first is that there is some
indication that star-galaxy separation is slightly less reliable in the
faintest half-magnitude slice of the APM survey than for brighter objects
(\refer{Maddox \etal\ 1996}).
The second is to show that the poor match between volumes sampled by
the bright redshift survey and the faint photometric survey does not
lead to a systematic bias in the estimated luminosity function.
In Figure~\ref{fig:magtest} we plot the LF estimated by counting APM 
galaxies to 20th, 19th, 18th and 17th magnitude respectively.
In each case, we see that the estimated LF is consistent with that
estimated from the full $b_J < 20.5$ sample.
Even cutting back the ``faint'' survey to $b_J < 17$ (actually slightly
shallower than the redshift survey), we still see evidence for a steep
faint end to the luminosity function.
One should note that the signal contributing to
the faint-end of the LF is coming
from decreasingly smaller volumes as one cuts back on the magnitude limit.
A galaxy of absolute magnitude $M_{B_J} = -15$ can only be seen to distances
of 93, 60, 39 and 25 $\hMpc$ respectively for galaxy samples cut to
$b_J = 20.0$, 19.0, 18.0 and 17.0.
The consistency of estimated faint-end slopes using galaxies over this
range of volumes argues strongly against the observed steep faint-end slope
being due to  a ``sampling'' or ``volume'' effect.

\subsection{Analysis of simulations} \label{sec:sims}

As a further test of our methods, we have generated two sets of 
mock-APM surveys
with known luminosity function and two-point correlation function,
thus allowing us to check that we indeed are measuring the ``true''
luminosity function.
The first set of simulations (``Sch'') has galaxy luminosities drawn from a 
flat faint-end Schechter function ($\alpha = -1$, $M^* = -19.5$).
The second set (``DP'') has a ``double power-law'' function of the form 
(\ref{eqn:dpschec})
with parameters $\alpha = -1$, $M^* = -19.5$, $\beta = -2.6$, $M_t =  -14.5$.
The idea is to check that our method for estimating $\phi(M)$ can
reliably distinguish between these two models.

For each model we generated five \refer{Soneira and Peebles (1978)} 
hierarchical clustering simulations, each containing 
2.4 million galaxies in the area of the APM survey.
From each of these mock APM surveys, we formed mock Stromlo-APM 
Redshift Surveys
by sampling one galaxy in 40 brighter than $b_J = 17.15$ at random.
Note that the sparser sampling rate compared with the real data (1 in 20)
is required to match the numbers of galaxies in the faint APM Galaxy Survey
and the brighter redshift survey sample.
This reflects the steep slope seen in the APM number counts by
Maddox et al (1990c) but not modeled in the simulations.
The mock redshift surveys contained on average 1648 galaxies each, and the
measured real space galaxy correlation function was well described on scales
1--20 $\hMpc$ by a power-law with parameters
$\gamma = 1.8$ and $r_0 = 4.9$ for the ``double power-law'' simulations,
and with $\gamma = 1.8$ and $r_0 = 5.1$ for the Schechter function simulations,
thus providing an excellent match to the observed real-space clustering 
of Stromlo-APM galaxies (Loveday \etal\ 1995).

We analysed the simulations in the same way as the real data;
counting galaxies in the mock-APM survey about centre galaxies in
the respective mock-Stromlo survey.
In Figure~\ref{fig:phi_sim}(a) we plot the mean $\phi(M)$ measured
from the five ``DP'' simulations as the symbols; the error bars
show the one-sigma scatter between the realizations.
These error bars are in good agreement with the errors estimated from 
equation~(\ref{eqn:phi_errs}).
Also shown are the individual estimates (dashed lines) and the true
luminosity function (continuous line).
We see that the mean estimated $\phi(M)$ is in reasonable agreement with
the true LF: if anything we tend to {\em underestimate} the
slope of the faint-end of the luminosity function.
We also see that estimates from the individual simulations can differ 
significantly from the true LF: two of the estimates show a {\em decrease} in
$\phi(M)$ faintward of $M \approx -17$.
This lack of robustness in the individual estimates of the LF is also 
apparent in 
Figure~\ref{fig:phi_sim}(b), where we plot the results from the ``Sch''
simulations.
Although the mean $\phi(M)$ is consistent with a flat faint-end,
and inconsistent with the double power-law function shown,
two of the estimates do show an apparent rise above the expected
flat faint-end slope.
Neither, however, maintains the rise as faint as $M = -12$.

Overall, although individual simulations can show a wide deviation from
the expected behaviour, it seems that we are more likely to underestimate
rather than overestimate the faint-end slope of the luminosity function,
and hence the space density of dwarf galaxies.
Our estimated errors are in good agreement with the scatter between
simulations.

\section{Conclusions and Discussion} \label{sec:concs}

We have seen that the number of dwarf ($M \gtrsim -15$) galaxies 
seen close in projection on the sky to $\sim L^*$ galaxies 
($-22 \lesssim M \lesssim -15$) is much
larger than expected for a flat faint-end Schechter function
and for the standard galaxy correlation function ($\gamma \approx 1.7$,
$r_0 \approx 5 \hMpc$).
We may thus infer that the space density of galaxies rises sharply
above a flat faint-end Schechter function for $M \gtrsim -15$
and/or that the clustering of dwarf galaxies around $\sim L^*$ galaxies is much
stronger than the auto-correlation function of $\sim L^*$ galaxies.
An extreme case of the latter explanation might be that dwarf galaxies
{\em only} exist close to $\sim L^*$ galaxies.

It is no easy matter to determine which (or both) of these explanations
is correct.
Since all of the $M \gtrsim -15$ dwarf galaxies must lie within
$y \lesssim 115 \hMpc$ to be visible in the APM survey, 
even if there is a generally higher space
density of dwarfs, they have a relatively minor ($\sim 20\%$) 
effect on total predicted number-counts in the survey.
However, one can place a lower limit on the space density of dwarf galaxies
by making the extreme assumption that they only exist close to $\sim L^*$ 
galaxies.
In fact, in Figure~\ref{fig:sigtest}, we see an excess
of dwarf galaxies up to a projected separation of at least $5 \hMpc$ from
$\sim L^*$ galaxies.
Thus a lower limit may be
estimated by assuming that they occur with the measured space density
only within $5 \hMpc$ of an $\sim L^*$ galaxy.
The lower limit on the average mean density of dwarf galaxies is then given 
by multiplying their measured space density
by the fraction of space within $5 \hMpc$ of an $\sim L^*$ galaxy.
%This ``filling factor'' is trivial to calculate for a random distribution
%of $\sim L^*$ galaxies, but of course the fact that galaxies are clustered
%will reduce this factor.
In order to estimate the filling factor of $\sim L^*$ galaxies in the 
local Universe,
we have generated another set of five Soneira-Peebles simulations
within a cubic volume
$100 \hMpc$ on a side and without applying a selection function.
These simulations contain 47,000 galaxies each, 
and thus their space density is
$\bar{n} = 4.7\ten{-2} \h3Mpcinv$, as measured for $\sim L^*$ galaxies in
the Stromlo-APM survey (Loveday \etal\ 1992).
Additionally, the two-point correlation function of galaxies in these
simulations is set to be well fit by a power-law with $\gamma = 1.82 \pm 0.06$
and $r_0 = 5.0 \pm 0.2$, in good agreement with the clustering measured 
by Loveday \etal\ (1995).
Given these simulations, it is then simple to perform a Monte-Carlo
calculation of the volume inside each cube within $5 \hMpc$ of an $\sim L^*$ 
galaxy.
We find that this factor lies in the range 0.57--0.62, 
ie. a randomly chosen point
in space has a 60\% chance of lying within $5 \hMpc$ of one or more 
$\sim L^*$ galaxies.
Now, integrating the luminosity function plotted in Figure~\ref{fig:phi}(b)
between $M = -15$ and $-12$ yields a measured density of dwarf galaxies
of $\bar{n} \approx 0.20 \h3Mpcinv$.
Correcting this by the extreme assumption that dwarf galaxies are only
found within $5 \hMpc$ of an $\sim L^*$ galaxy results in a {\em lower limit}
on the space density of dwarf galaxies of $\bar{n} \approx 0.12 \h3Mpcinv$.
This is a factor of two higher than the density 
$\bar{n} \approx 0.058 \h3Mpcinv$ inferred from the $\alpha = -1.11$ Schechter
function fit by Loveday \etal\ (1992).

In fact, the true space density of dwarf galaxies is likely to be
significantly higher than this, for a number of reasons.
First, our estimator assumes that galaxy clustering is independent of 
luminosity, whereas we know that sub-$L^*$ galaxies are less strongly clustered
than more luminous galaxies (eg. Loveday \etal\ 1995).
If this luminosity segregation extends to dwarf galaxies, then we will have
underestimated their space density.
Second, analysis of simulations (\S\ref{sec:sims}) shows that our estimator
tends to underestimate the faint-end slope of $\phi(M)$ slightly,
possibly due to a Malmquist-type bias.
Third, as discussed by numerous authors, most galaxy surveys
are likely to be missing a substantial fraction of low surface brightness
galaxies, many of which will be dwarfs.
For example, \refer{Sprayberry \etal\ (1997)} find a pronounced upturn in the
luminosity function for their sample of low surface-brightness galaxies.
Thus we regard our above estimate of the space density of dwarf galaxies,
$\bar{n} \approx 0.12 \h3Mpcinv$, as a {\em lower limit} on the true value.

A high space density of dwarf galaxies, assuming that they are predominantly
late-type, blue galaxies, which suffer smaller $K$-correction dimming
than redder, early-type galaxies, provides a natural explanation for
the steep observed number counts of faint galaxies.
Evidence for a large contribution from late-type galaxies to the faint 
galaxy counts in the HST Medium Deep Survey has been presented by
\refer{Driver, Windhorst and Griffiths (1995)}.
Zucca \etal\ (1997) have found that the luminosity function for emission-line
galaxies (ELGs) is significantly steeper at the faint end than the LF
for non-ELGs, also supporting the hypothesis that the faint-end of
the galaxy LF is dominated by late-type galaxies.
Gronwall and Koo (1995) were able to match observations of galaxy number counts
in the $K$, $R$ and $B_J$ bands, as well as colour and redshift distributions,
for a mild evolution model by assuming that the faint end of the galaxy 
luminosity function is dominated by blue galaxies ($B - V \le 0.6$),
and rises significantly above a Schechter function with flat faint-end slope.
We plot the Gronwall \& Koo model total luminosity function in
Figure~\ref{fig:phi}(b), and see remarkably good agreement with
our observations.
Our results thus support the model of Gronwall \& Koo; 
there is no need to invoke
exotic forms of galaxy evolution to explain observed galaxy number counts
at faint magnitudes.

\acknowledgments

It is a pleasure to thank my colleagues George Efstathiou, Steve Maddox,
Bruce Peterson and Will Sutherland, who made the APM and Stromlo-APM
surveys possible.
The simulations used in this paper were carried out using computing
facilities at the Fermi National Accelerator Laboratory.
The referee, Chris Impey, is thanked for his comments which have
(hopefully) helped make these results more convincing.

\clearpage
\section*{Figure Captions}

\figcaption{Plots of the ratio of APM galaxies counted around centre
galaxies ($N_{\rm gal}$) over the number of APM galaxies in randomly
placed control fields ($N_{\rm bgr}$) of the same solid angle.
The solid histogram shows the ratio $N_{\rm gal}/N_{\rm bgr}$ and the
dotted histogram shows the inverse $N_{\rm bgr}/N_{\rm gal}$.
APM galaxies were counted to a $b_J$ magnitude limit of (a) 20.0, (b) 19.0,
(c) 18.0 and (d) 17.0.
\label{fig:bgrtest}}

\figcaption{Estimates of the galaxy luminosity function $\phi(M)$.
(a) After just one iteration.  
The continuous line shows the
earlier Schechter function fit to $\phi(M)$ using the redshift survey 
galaxies alone (\refer{Loveday \etal\ 1992}); its extrapolation
to lower luminosities is shown by the dotted line.
(b) After ten iterations, by which time the solution has converged.
The smooth curve shows a ``double power-law'' Schechter fit
(eqn.~\ref{eqn:dpschec}) and the histogram shows the Gronwall \& Koo (1995)
model.
\label{fig:phi}}

\figcaption{Estimates of the galaxy luminosity function $\phi(M)$
obtained by counting APM galaxies to a projected separation 
$\sigma \le 2 \hMpc$ 
(solid symbols) and $\sigma \le 5 \hMpc$ (open symbols).
The smooth curve is the same ``double power-law'' Schechter fit
shown in \protect{Figure~\ref{fig:phi}b}.
\label{fig:sigtest}}

\figcaption{Estimates of the galaxy luminosity function $\phi(M)$
obtained by counting APM galaxies to a $b_J$ magnitude limit of 
(a) 20.0, (b) 19.0, (c) 18.0 and (d) 17.0.
The smooth curve is the same ``double power-law'' Schechter fit
shown in \protect{Figure~\ref{fig:phi}b}.
\label{fig:magtest}}

\figcaption{Analysis of simulations.
The dashed lines show the individual realizations, the symbols with error bars 
show the mean and 1-sigma fluctuations between them.
The continuous line shows the ``double power-law'' LF used in the ``DP'' 
simulations.
(a) Results from the ``double power-law'' simulations (``DP'').
(b) Results from Schechter function simulations (``Sch'').
\label{fig:phi_sim}}

%\end{document}
% comment out above line to include the figures

% Finally, include the figures
%\epsfverbosetrue
\epsfxsize=\textwidth
\setcounter{figure}{0}

% lum:hist_plot.f, hist_m20.dat, hist_m19.dat, hist_m18.dat, hist_m17.dat
\begin{figure}[p]
\plotone{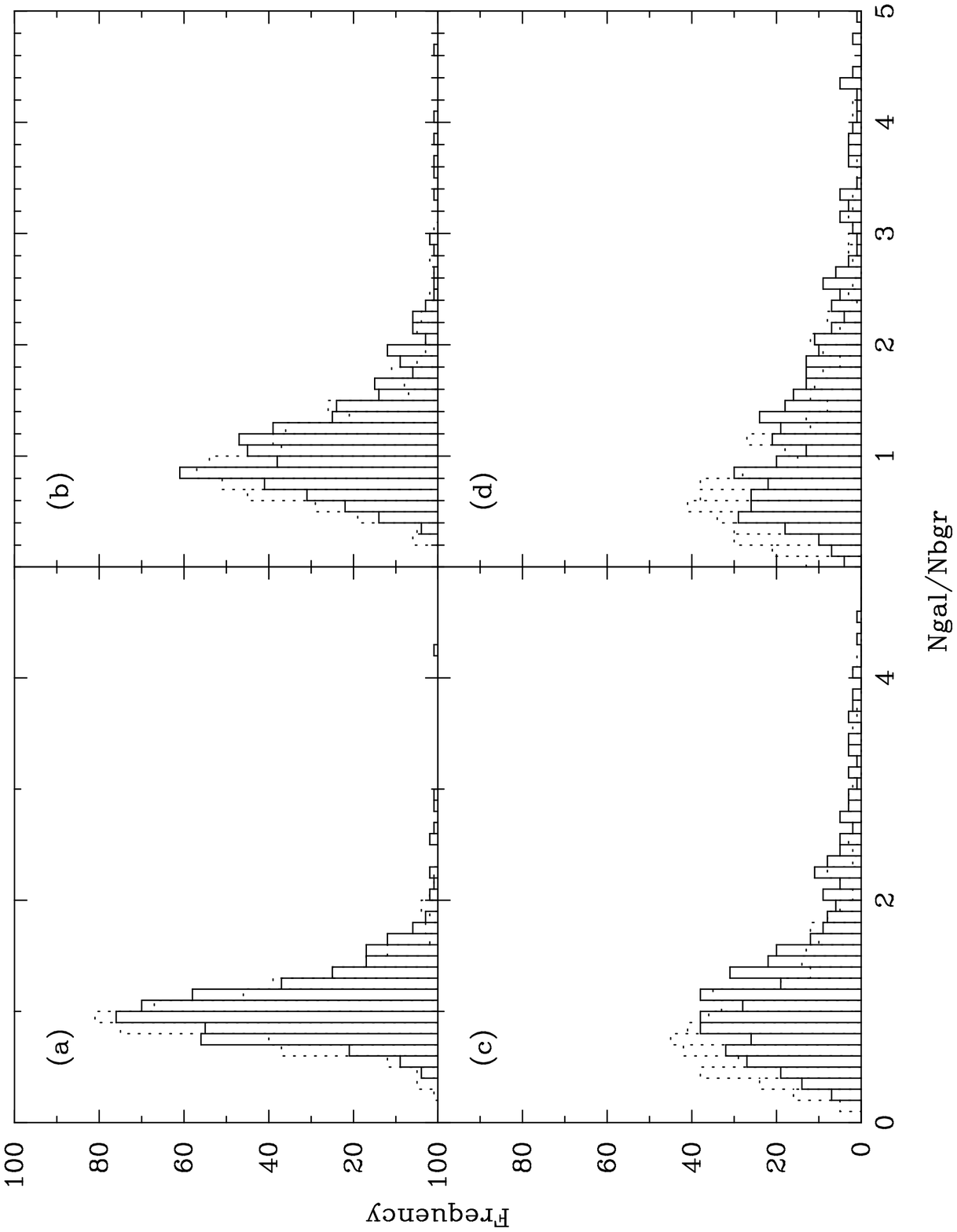}
\caption{\relax}
\end{figure}

% lum:lumx_plot.f, lumdat:phi_1st_itr.dat, phi_20.5_1.dat
\begin{figure}[p]
\plotone{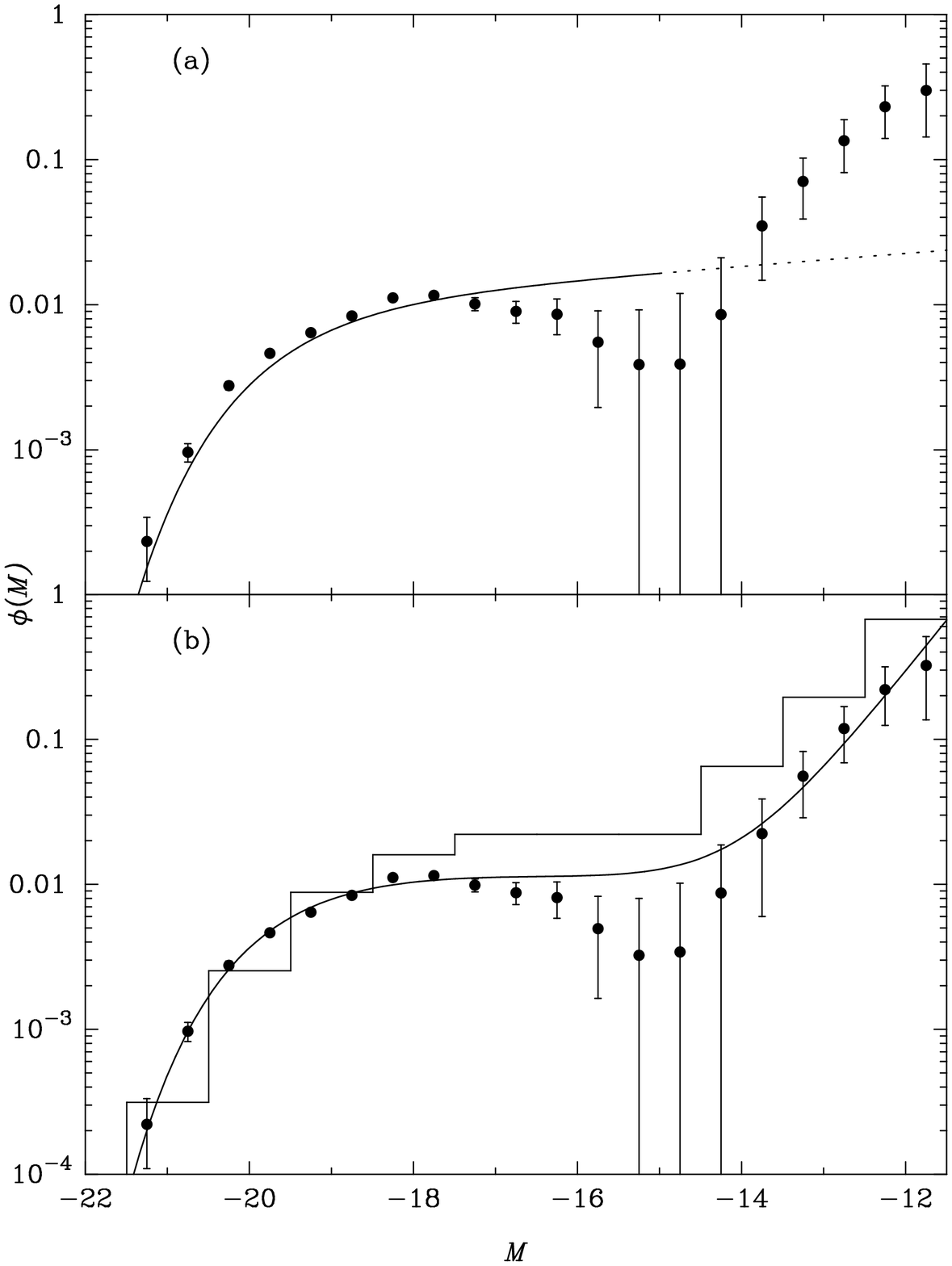}
\caption{\relax}
\end{figure}

% lum:lumx_plot.f, lumdat:phi_20.5_2.dat, phi_20.5_5.dat
\begin{figure}[p]
\plotone{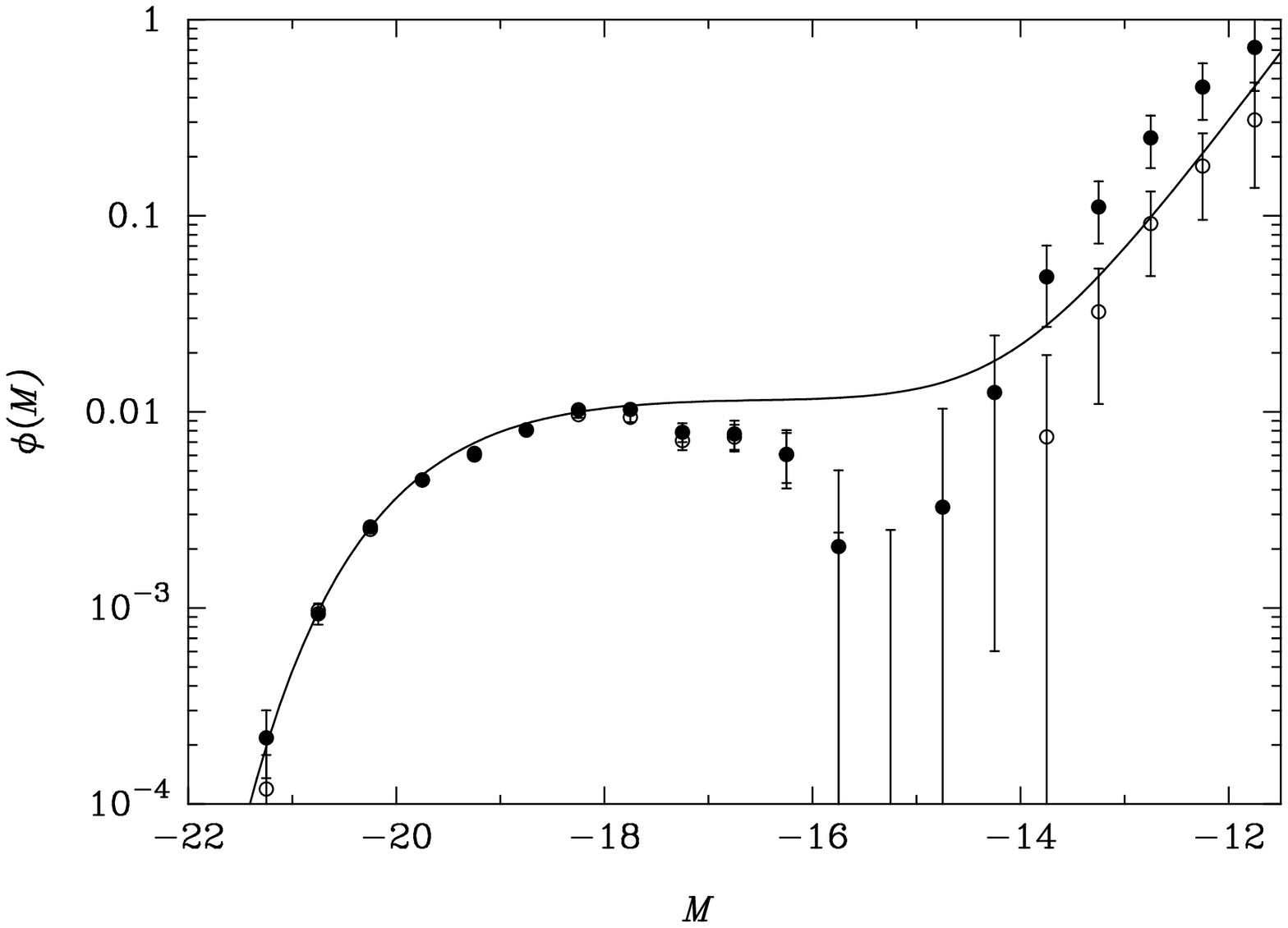}
\caption{\relax}
\end{figure}

% lum:lumx_plot.f, phi_20.0_1.dat, phi_19.0_1.dat, phi_18.0_1.dat, 
% phi_17.0_1.dat, 
\begin{figure}[p]
\plotone{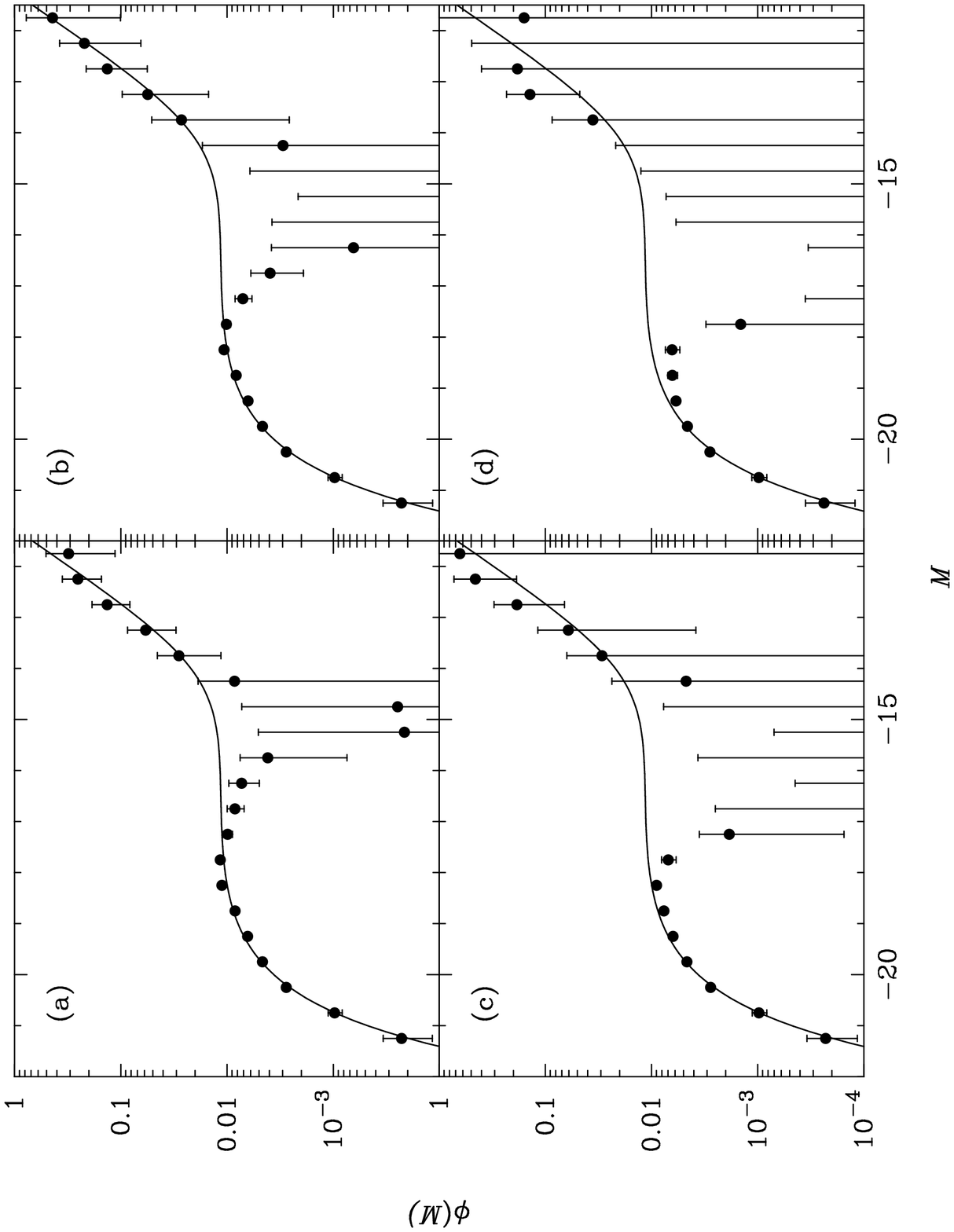}
\caption{\relax}
\end{figure}

% lum:lumx_plot.f, simdat:phi_sim_dpav.dat, phi_sim_dpN.dat, phi_sim_schav.dat,
% phi_sim_schN.dat
\begin{figure}[p]
\plotone{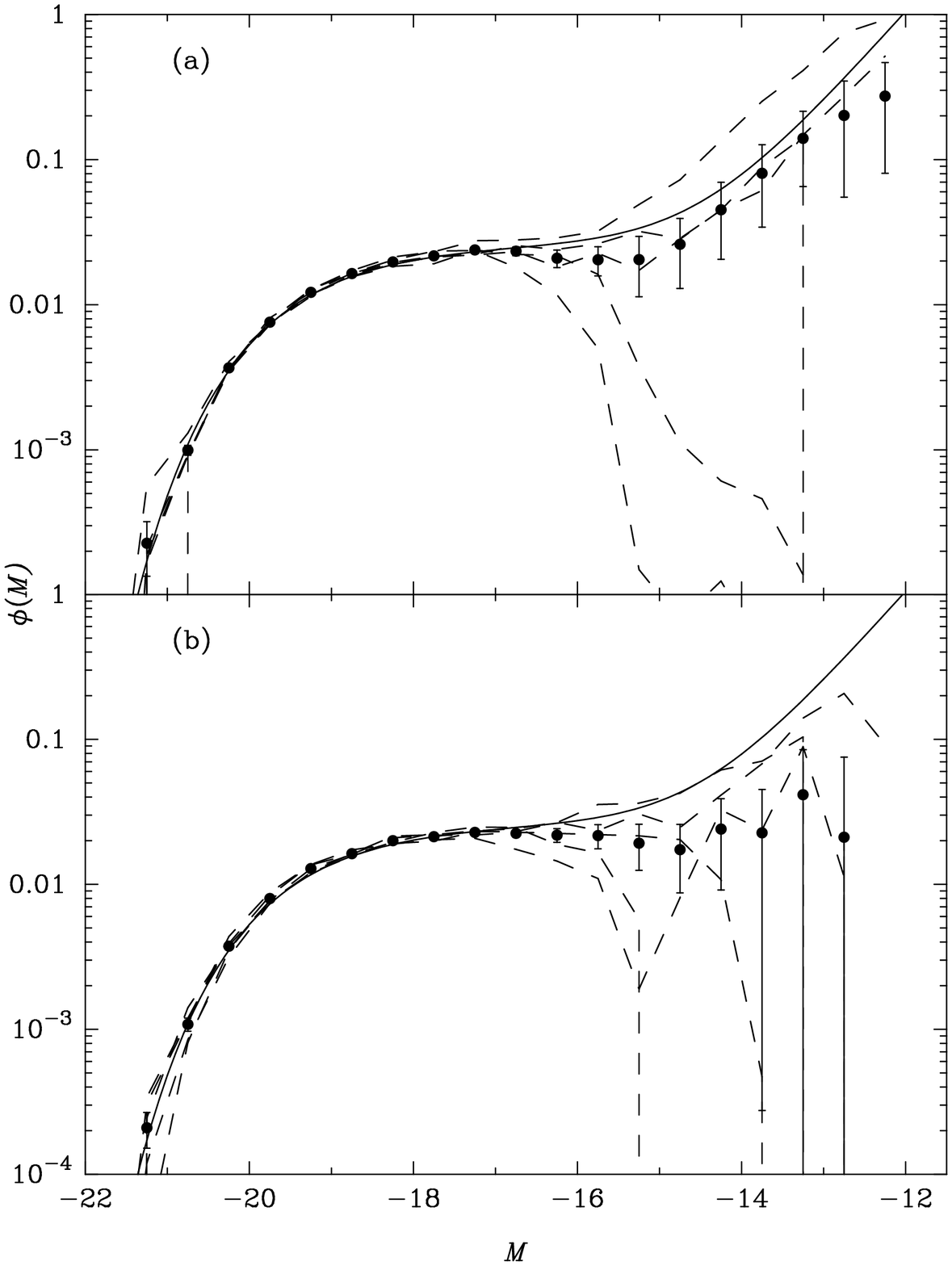}
\caption{\relax}
\end{figure}

\end{document}